\documentclass[prl,floatfix,twocolumn,showpacs,amsmath,amssymb]{revtex4}
\usepackage{graphicx,color}
\usepackage{dcolumn}
\usepackage{bm}

\begin{document}

\title{Magneto-elastic instability in Molecular Antiferromagnetic Rings}

\author{Leonardo Spanu$^{1}$ and Alberto Parola$^{2}$}
\affiliation{ ${^1}$ INFM and Dipartimento di
Fisica ``A.Volta", Universit\`a di Pavia, I-27100 Pavia, Italy\\
${^2}$ INFM and Dipartimento di
Fisica e Matematica, Universit\`a dell'Insubria, I-22100 Como, Italy 
}
 
\date{\today}

\begin{abstract}
Lattice stability in a model of antiferromagnetic ring
coupled to adiabatic phonons is investigated for different 
values of the spin and numbers of magnetic sites. 
The magneto-elastic transition is shown to be heavily affected by the spin
value, displaying a qualitative difference in the nature of the 
instability for spin one-half. Among the different synthesized materials, 
Cu$_8$ seems to be the best candidate to observe lattice dimerization 
in these systems. Our analysis excludes stable lattice distortions 
in higher spin rings.  The effects of thermal fluctuations are 
studied in the Cu$_8$ model, where a characteristic crossover 
temperature is estimated. 
\end{abstract}


\maketitle



Magnetic molecular rings (MMR),with nearest neighbors anti-ferromagnetic (AF) 
interactions, are ideal materials for exploring the crossover from microscopic 
magnetism to collective effects in low dimensional systems.
Rings with different value of the spin $S$ and  number $N$ of magnetic 
sites have been synthesized in recent years ~\cite{caneschi}. 
They give the opportunity to understand how  the static and dynamical 
properties of the material are affected by a change in the spin value 
or a change in the number of magnetic sites.
Considerable attention has been devoted to the observation of resonant
tunneling of the magnetization in single-molecule
superparamagnets ~\cite{wesd} and 
of quantum steps of magnetization in molecular AF rings
~\cite{loss}~\cite{fe6chem}. 


Ring shaped molecular clusters, such as
Fe$_6$~\cite{fe6chem},Fe$_{10}$~\cite{fe10},Cr$_8$~\cite{cr8chem},Cu$_8$
~\cite{cu8chem}, have been studied by several techniques, such as 
nuclear magnetic resonance (NMR)~\cite{NMRfe6},~\cite{cu8}, calorimetric
measurements, X-rays and inelastic neutron scattering (INS).
Recently, INS experiments performed on Cr$_8$ molecular rings ~\cite{cr8}
have been interpreted on the basis of a distorted ring geometry at low 
temperature, in contrast to the centrosymmetric structure suggested 
by X-ray diffraction at room temperature.
In the Cu$_8$ rings, the Cu NQR spectrum shows four 
structurally non equivalent Cu ions ~\cite{cu8}: An open question is whether 
the reduced symmetry is related to the appearance of dimerization. 
Heat capacity and magnetic torque measurements in
magnetic field may also suggest the occurrence of a dimerized phase 
in Fe$_6$:Li at $T=1$ K~\cite{affronte}.

The Spin-Peierls (SP) transition is a magneto-elastic instability which 
occurs in a spin chain coupled with the lattice. It occurs when a distorted 
phase characterized by a lattice dimerization is stabilized below a critical 
temperature $T_c$ due to the gain in magnetic energy.
Such a transition was first predicted to occur in the (infinite) 
$S=1/2$ Heisenberg AF chain ~\cite{primo} and it was indeed observed 
experimentally in the quasi one dimensional $S=1/2$ compound 
CuGeO$_3$ ~\cite{hase}.
A number of  results have been obtained both analytically ~\cite{teo1} 
and numerically ~\cite{teo2} for the spin one-half Heisenberg 
chain coupled to the phonons.  Few investigations have been performed in 
the case of finite rings and/or higher spin~\cite{guo}. On general grounds we observe that 
in the thermodynamic limit, the dimer susceptibility of the Heisenberg chain
is known to diverge as $|\pi-q|^{-1}$ (with logarithmic corrections) for 
wave vectors close to $q=\pi$ and {\sl arbitrary} semi-integer spin $S$ 
\cite{haldane}, leading to SP instability at $T=0$. 
The situation is different in a finite-size system or in chains with
integer spin $S$, where the singlet gap in the excitation spectrum gives a finite 
susceptibility also in the $T\rightarrow 0$ limit. 
Therefore we expect that 
in MMR the lattice distortions will be stabilized only for a sufficiently 
small elastic constant. The competing role of finite size effect and of thermal 
fluctuations in these systems has not been clarified yet. 

In this Letter we first analyze the stability of a mesoscopic magnetic ring with 
respect to lattice distortions and then we examine the finite temperature
effects in these systems. 
Using Lanczos Diagonalization Techniques and Spin Wave Theory (SWT) we 
show how the magneto-elastic transition is affected by a change of the 
spin value. Surprisingly, a {\sl qualitative} difference between the behavior of 
$S=1/2$ and higher spin rings is found, strongly suggesting the absence of stable
lattice distortion in the latter case. Among the synthesized MMR, Cu$_8$ 
seems to be the only candidate to display a magneto-elastic instability
for realistic values of the phonon coupling. 
An estimate of the crossover temperature associated to this `transition" is also given.

We  focus on rings with an even number of sites, in which the 
AF superexchange coupling leads to a singlet ground state. 
We investigate the simplest model Hamiltonian containing both 
AF interaction and spin-lattice coupling

\begin{equation}
\label{hami}
{\mathcal{H}} = J \sum_{j=1}^{N} [1+\alpha \delta_{j}]\vec{S}_{j}\cdot
\vec{S}_{j+1} + \frac{K}{2}\sum_{j=1}^{N}\delta_{j}^{2}
\end{equation}
where $K$ is the spring constant, $\delta_i$ is the distortion of the bond 
between site $i$ and $i+1$, $\alpha$ is the spin-lattice coupling
and the last term is the elastic energy. Periodic boundary conditions are 
understood ($S_{N+1}=S_1$). Lattice distortions are treated in 
the adiabatic approximation, i.e. $\delta_i$ are c-numbers. 
Due to the uncertainties in the modeling of the microscopic mechanisms 
leading to the spin-phonon coupling in these materials, harmonic approximation
has been employed in the form of Hamiltonian (\ref{hami}), thereby limiting 
its validity to the $\alpha|\delta_{i}| \ll 1$ case.
We did not include in the Hamiltonian other terms,
usually present in the microscopic description of MMR ~\cite{cr8},
which do not directly couple with the lattice, such as dipolar interactions 
or spin anisotropies. The invariance of the Hamiltonian under the 
rescaling $\alpha \to \lambda\alpha$, $K \to \lambda^2 K$ and 
$\delta_i \rightarrow \delta_{i}/\lambda$ allows to restrict the 
parameter space by setting $\alpha=1$ without loss of generality.
The further constraint, due to periodic boundary conditions, 
$\sum_i\delta_i=0$ is also understood. By relaxing this constraint, the 
numerical results remain qualitatively unchanged. 

The ground state of Hamiltonian (\ref{hami}) is obtained in two steps
\cite{feigun}:
first, by Lanczos technique, we determine the lowest energy eigenvalue of
(\ref{hami}) at fixed $\{\delta_i\}$. The bond distortions
are then updated in order to comply with the equilibrium condition
obtained by use of the Hellman-Feynman theorem:
\begin{equation}
\label{equi}
K \delta_j - J  \langle \vec{S}_{j}\cdot \vec{S}_{j+1} \rangle_{\delta} 
-\frac{J}{N} \sum_{i=1}^{N} \langle \vec{S}_{j}\vec{S}_{j+1} \rangle_{\delta}=0
\end{equation}
where $\langle ... \rangle_{\delta}$ is the ground state expectation value, 
in presence of distortion. 
No lattice symmetry has been assumed, in order to investigate all 
the possible lattice equilibrium configurations.
In all cases we examined, we found either an undistorted configuration 
or a dimerized ring, i.e. $\delta_i=(-1)^i \delta_0$.
No other periodicity has been observed, as already known for $S=1/2$ 
~\cite{lieb}.

\begin{figure}
\vspace{2mm}
\includegraphics[width=0.43\textwidth]{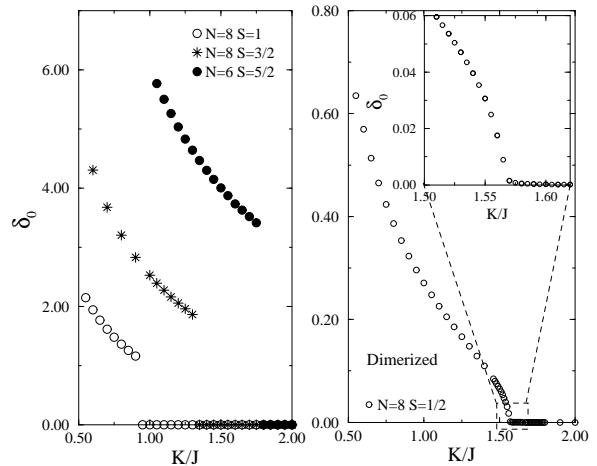}
\caption{\label{fig1}
Stability diagram for different sizes $N$ and spin $S$, corresponding to
synthesized materials. Spring constant $K$ in units of $J$. 
Left panel: $N=6$ $S=5/2$ (Fe$_6$); $N=8$ $S=3/2$ (Cr$_8$)
and $N=8$ $S=1$. 
Right panel: $N=8$, $S=1/2$ (Cu$_8$).
Inset: blow-up of the Cu$_8$ stability diagram for $K\sim K_c$
}
\end{figure}

The numerical results concerning the stability of the undistorted phase 
are reported in Fig. \ref{fig1}. The distortion amplitude is plotted 
as a function of the elastic constant, for different values of the spin. 
All models we examined show a critical constant $K_c$ beyond which the 
symmetric phase is unstable.
The results clearly show that, while in the spin one-half case the 
transition is continuous, for larger spin the distortion 
amplitude $\delta_{0}$ suddenly jumps from zero to a finite value 
(even larger than the lattice spacing) when the spring constant is reduced. 
Although such a large distortion falls outside the range of validity of the 
model, the results clearly point toward a first order transition in 
the $S>1/2$ case. No qualitative differences between integer and 
semi-integer spins can be observed in these small rings. 

\begin{figure}
\vspace{2mm} 
\includegraphics[width=0.33\textwidth]{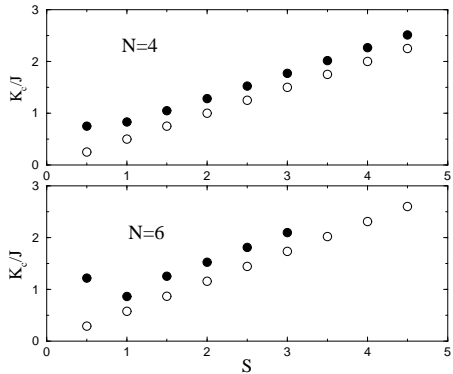}
\caption{\label{fig2}
Scaling relation for $K_c$ in units of $J$ 
as a function of the spin $S$, for $N=4$ 
(top panel) and $N=6$ (bottom panel). Black
filled symbols are Lanczos data, empty symbols spin-wave results }
\end{figure}

In order to understand the origin of the different behavior of 
the $S=1/2$ case, we have performed
Lanczos diagonalizations fixing the dimerization amplitude $\delta_0$ and
we computed the ground state energy of the dimerized Heisenberg model 
$E_H(\delta_0)$. In terms of this quantity the equilibrium condition reads
\begin{equation}
\label{equi0}
\frac{d}{d \delta_0} \left[ E_H(\delta_0)+\frac{N}{2}K
\delta_0^2 \right]=0 
\end{equation}
$\delta_0=0$ is always a solution, $E_H(\delta_0)$ being analytic in 
$\delta_0^2$. Generally, lattice distortions lower the magnetic energy 
$E_H$ for any $S$, making the undistorted phase stable only for $K>K_c$, with
\begin{equation}
K_c=-\frac{1}{N} E_H^{\prime\prime}(\delta_0) \Big\vert_{\delta_0=0} 
\label{kc}
\end{equation}
where the prime denotes derivative with respect to the dimerization amplitude.
The second derivative of the ground state energy (i.e. the dimer susceptibility)
at r.h.s. of Eq. (\ref{kc}) can be evaluated numerically for several 
$N$ and $S$.  The resulting critical coupling $K_c$ are shown in Fig. 
\ref{fig2} suggesting an approximate linear scaling with $S$.
Note that the natural energy scale 
of the problem is set by the classical magnetic term $\sim JS^2$. Therefore, the
{\it dimensionless} critical spring constant $K_c/JS^2$ actually scales
as $1/S$ and vanishes in the classical limit $S\to\infty$. 
Following the standard Landau-Ginzburg approach, the equilibrium 
distortion amplitude is predicted to continuously evolve from zero to
finite values as $K$ is reduced below $K_c$.
This expectation rests on the assumption of a {\it positive}
fourth order term in the expansion of $E_H(\delta_0)$ about $\delta_0=0$.
We checked this condition by computing the fourth derivative
of $E_H$ at $\delta_0=0$ which turns out to be positive 
only in the $S=1/2$ case, as shown in Table \ref{tab}. 
Therefore for $S>1/2$, the transition is
indeed first order as suggested by the numerical results of Fig. \ref{fig1}.
This numerical finding is confirmed by studying the $S\to\infty$ limit of
our model (\ref{hami}). For a dimerized ring, the classical 
ground state energy $E_H^{cl}=- J \sum_{i} (1+(-1)^{-i}\delta_{0})S^{2}=-NJS^2$ 
becomes independent of the distortion amplitude $\delta_{0}$.
It follows that, classically, the distortion does not decrease the magnetic
energy contribution and hence a dimerized geometry is not favored. Only 
quantum fluctuations lift the degeneracy in 
$\delta_0$. In order to estimate the effects of quantum fluctuations on
the ground state energy we made use of lowest order spin-wave theory (SWT)
for finite size chains \cite{zhong} which is free from singularities 
even in 1D.  By applying SWT to the dimerized Heisenberg ring, 
we obtain the $O(S)$ expression for $E_H(\delta_0)$:
\begin{equation}
\label{ene0} 
E_H(\delta_0)=-NJ S(S+1) + 2S J \sqrt{1-\delta_0^2} \cot\frac{\pi}{N}
\end{equation}
leading to the result:
\begin{equation}
\label{equi3}
K_c =-\frac{1}{N}\frac{d E_H}{d \delta_0^2}\Big\vert_{\delta_0=0}
    =\frac{2S J\cot \frac{\pi}{N}}{N}
\end{equation}
This explains the linear scaling observed in the
numerical results. Direct evaluation of the fourth derivative of 
$E_H(\delta_0)$ from Eq. (\ref{ene0}) confirms the first order transition
in the classical limit $S\to\infty$ as shown in Table \ref{tab}.
The continuous transition for spin
one-half rings is therefore exclusively due to the enhancement of
quantum fluctuations and is then a genuine quantum effect which goes
beyond semi-classical treatments. 

\begin{figure}  
\vspace{2mm}
\includegraphics[angle=-90,width=0.35 \textwidth]{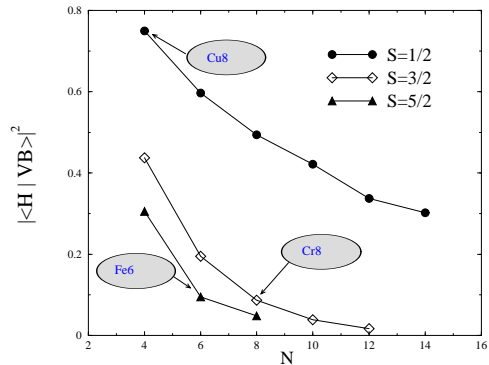}
\caption{\label{fig3}
Overlap between the undistorted Heisenberg ground state $|H>$ and the VB state
$|VB>$ for different spins as a function of $N$}
\end{figure}

Some further physical insight on the difference between spin one-half and $S>1/2$
can be gained by investigating the ``proximity" between the undistorted 
ground state and the valence bond (VB) state, corresponding to $\delta_0=1$,
which mimics the dimerized phase in the strong distortion limit. 
By calculating the overlap between these
two states shown in Fig. \ref{fig3} we appreciate a noticeable 
dependence on the spin value, the overlap getting smaller and smaller 
for increasing $S$ and being significant only for $S=1/2$. 
Such a decrease in the overlap suggests that a continuous transition 
between the undistorted and the distorted phase becomes less and less
favored for spin larger than one-half at any size $N$.

\begin{table} [b]
\begin{tabular}{|c|c|c|c|c|c|c|}\hline
     &$S=1/2$ &$S=1$ &$S=3/2$ &$S=2$ &$S=5/2$ &$S=\infty$  \\
\hline
$N=4$&$54.0$ &$-18.9$ &$-9.7$ &$-8.2$ &$-7.6$ &$-6.0$  \\
\hline
$N=6$&$423.3$ &$-72.1$ &$-12.3$ &$-14.7$ &$-13.3$ &$-10.4$  \\
\hline
$N=8$&$1589.8$ &$-152.2$ &$-5.5$ &  &   &$-14.5$   \\
\hline
\end{tabular}
\caption{Fourth derivative of the ground state energy with respect to
the dimerization parameter $\delta_0$ for different sizes $N$ and spins $S$.
Values are divided by $S$. The $S=\infty$ results are obtained by use of spin
wave theory, Eq. (\ref{ene0}).
}
\label{tab}
\end{table}

\begin{figure}  
\vspace{2mm}
\includegraphics[angle=0,width=0.3 \textwidth]{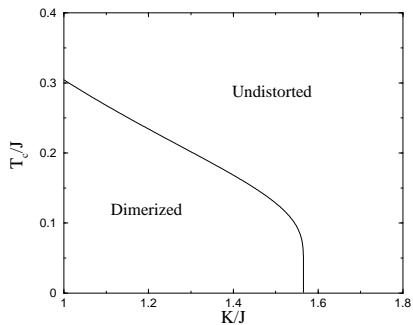}
\caption{\label{fig4}
Crossover temperature as a function of the elastic coupling for a $N=8$,
$S=1/2$ MMR}
\end{figure}

Finally we comment on the stability of the distorted phase against
thermal fluctuations. 
The coupling of the spins with the lattice (considered here
as embedded in the tree dimensional space)
allows for a magneto-elastic transition at finite-temperature,  
driven by the one-dimensional magnetic fluctuations.     
We concentrate on the the behavior of 
$S=1/2$ rings for only in these cases the ground state may show stable
lattice distortions. In finite size systems no sharp transitions 
may be observed at finite temperature and only {\sl crossover} temperatures
can be defined. However, a sharp change at some $T_c(K)$ 
is expected in the distortion distributions for $K<K_c$: The low temperature
 bimodal distribution peaked at $\pm\delta_0$ turns into a Gaussian-like
probability centered around $\delta_0=0$ for temperatures larger than the
``mean field" critical temperature $T_c(K)$. By employing exact diagonalization
in the full Hilbert space of the Hamiltonian (\ref{hami}) we computed 
the partition function and the finite temperature dimer susceptibility 
of the model thereby obtaining the critical temperature shown 
in Fig. \ref{fig4}. 
Note the steep increase in $T_c(K)$ close to the
transition, implying that temperatures higher than $\sim 0.1\,J$ are
always required to destabilize the lattice distortion.

Now we are ready to  apply our results to real materials and discuss the 
possible occurrence of a dimerized phase in magnetic molecular rings
The results of Fig. \ref{fig1} strongly suggest that lattice distortions
cannot be observed in synthesized magnetic rings with $S>1/2$. In these cases
small distortions are never stabilized even at $T=0$, while the 
occurrence of a strongly first order transition would 
probably disrupt the ring structure in the
absence of a sufficiently hard elastic constant $K$.
On the contrary, as show in Figs.(\ref{fig1},\ref{fig4}), in the case of Cu$_8$ 
($N=8$ magnetic sites and $S=1/2$) a stable dimerized phase is instead 
possible at low temperature. 
According to Ref. ~\cite{cu8}, the Cu$_8$ ring has a 
coupling of about $J=0.1$ eV, the exchange interaction being due to Cu-O-Cu 
bridge as usual in copper-oxide compounds. Assuming a typical value for
$\alpha\sim 10$, we find that small lattice distortions can be stabilized
by a spring constant attaining the physically reasonable value 
$K\simeq 0.1 J\alpha^2\sim 1$ eV. 
Instead, in 
iron and chromium rings the magnetic coupling is much weaker 
($J\sim 1.5$ meV in  Cr$_8$ ~\cite{cr8chem}~\cite{cr8} and $J\sim 2.4$ meV in Fe$_6$
~\cite{fe6chem}~\cite{NMRfe6})
pushing lattice instabilities to considerably smaller value of $K$.

In conclusion, using SWT and Lanczos diagonalizations, we have investigated
the properties of a generic AF magnetic ring, with spins adiabatically coupled to lattice 
distortions. We have emphasized the role of the spin in the occurrence of 
magneto-elastic instabilities, showing that the order of the transition 
strongly depends on the value of $S$. In order to assess the 
accuracy of the adiabatic approximation employed here, experimental 
studies on the characteristic frequencies of the phonon modes will be
necessary. However, we believe that this test is not crucial for the 
present investigation because lattice instabilities are classical 
phenomena whose occurrence can be reliably estimated also in the adiabatic limit. 
Our analysis suggests that magneto-elastic coupling cannot be invoked
for the interpretation of NMR and NQR data in MMR with spin different from
one-half~\cite{affronte},
while confirming this possibility in the case of Cu$_8$~\cite{hase,cu8}.
An investigation of the effects of thermal fluctuations on the stability
of the distorted phase has also been performed for the latter case.
X ray scattering at low temperature and NMR spectra analysis
will be able to determine whether MMR do indeed display lattice 
distortions in the ground state.

L.S. thanks F. Borsa, A. Lascialfari and Y. Furukawa
for useful discussions and for showing and explaining their experimental
results. We also thank F. Becca and N. Masciocchi for valuable comments. 
L.S. acknowledges kind hospitality at the Department of Physics of
the university of Milano.

\end{document}